\documentstyle[aps,preprint]{revtex}
\begin{document}
\draft

\title{Renormalized vs.\ unrenormalized perturbation-theoretical
approaches to the Mott transition} 
\author{Ekkehard Lange}
\address{Serin Physics Laboratory,
Rutgers University,
136 Frelinghuysen Road,
Piscataway, New Jersey 08854}
\maketitle

\begin{abstract}

I employ an exactly soluble toy model to investigate why
unrenormalized perturbation theory works better than fully 
self-consistent approaches in describing the correlation-driven
metal-insulator transition.

\end{abstract}


Theoretical works on the Mott transition \cite{Mott:1961} have mainly
focused on the Hubbard model of strongly correlated electrons
\cite{Hubbard:1963}. Its main feature is the competition between
itineracy and an on-site repulsion $U$. At half filling and zero
temperature, the Hubbard model exhibits a metal-insulator transition 
at some critical value $U_c$. At the moment, the only unified
framework for describing the various features and phases associated
with this Mott transition is provided by the dynamical mean-field
theory \cite{Georges:1996}, which becomes exact in the limit of
infinite dimensions introduced by Metzner and Vollhardt
\cite{Metzner:1989a}. However, the dynamical mean-field equations
cannot be solved exactly, and various approximate numerical schemes
have been employed to study their intricate structure. An important
approximation for describing the half-filled Hubbard model and the
Mott transition is the so-called iterated perturbation theory (IPT). 
This approximation succeeds to simultaneously capture the
quasiparticle resonance and the incoherent lower and upper Hubbard
bands. Partly based on the IPT, Zhang, Rozenberg, and Kotliar put
forward a scenario for the Mott transition, in which the width of the
quasiparticle resonance vanishes linearly in $U-U_c$ while its height
remains unchanged, as the transition point $U_c$ is approached from
below \cite{Zhang:1993}. This scenario was directly confirmed by the
projective self-consistent technique of Moeller {\it et al.}
\cite{Moeller:1995}. Yet, the IPT is based on second-order
perturbation theory in $U$ and might therefore be suspected to be
inferior to perturbation-theoretical schemes that are either fully
self-consistent or involve larger classes of diagrams. Self-consistent
perturbation expansions have the advantage of conserving exact
symmetries and their respective conservation laws \cite{Baym:1961}. 
Therefore, some authors have adopted a self-consistent approach from
the outset \cite{Mueller-Hartmann:1989b,Menge:1991}. Others, by
contrast, have pointed out early on that fully self-consistent
approaches seem to work worse than ordinary perturbation theory
\cite{Georges:1992a}. They relied on weak-coupling studies of the
half-filled single-impurity Anderson model
\cite{Yosida:1970,Yosida:1975,Yamada:1975}. Meanwhile, there is a
large body of numerical evidence, that, at least not too close to the
Mott transition, IPT is quantitatively accurate
\cite{Zhang:1993,Kajueter:1996b}. But it is not fully understood why
leading-order perturbation theory works so well, whereas seemingly
more elaborate self-consistent approaches notoriously fail to properly
describe the insulating phase and their precursor effects.

It is the purpose of this paper, to shed some light on this issue by
considering an exactly solvable toy model that crudely captures some
features of the Mott transition. This model is defined by the
following Hamiltonian:
\begin{equation}
 \hat{H}=U(\hat{n}_{\uparrow}-\frac{1}{2})
          (\hat{n}_{\downarrow}-\frac{1}{2})
	+V\sum_{\sigma}(f^+_{\sigma}c_{\sigma}+c^+_{\sigma}f_{\sigma}),
\label{toy-model}
\end{equation}
where $\hat{n}_{\sigma}\equiv c^+_{\sigma}c_{\sigma}$. It describes a
correlated $c$ orbital hybridizing with an $f$ orbital at zero
energy. At half filling and zero temperature, particle-hole symmetry
guarantees that the Green's function depends on three independent
parameters only,
\begin{eqnarray}
 G(z)&=&\sum_{j=1}^2\left(\frac{a_j}{z-\epsilon_j}
			 +\frac{a_j}{z+\epsilon_j}\right),
\label{green}\\
 \epsilon_1&=&\frac{1}{4}\left(\sqrt{U^2+64V^2}
			      -\sqrt{U^2+16V^2}\right),
\label{eps1}\\
 \epsilon_2&=&\frac{1}{4}\left(\sqrt{U^2+64V^2}
			      +\sqrt{U^2+16V^2}\right),
\label{eps2}\\
 a_1&=&\frac{1}{4}\left(1-\frac{U^2-32V^2}
		{\sqrt{(U^2+64V^2)(U^2+16V^2)}}\right),
\label{a1}
\end{eqnarray}
where $z$ is a complex frequency and $a_2=1/2-a_1$. At $V=0$, the
Green's function has two poles of equal weight at $\pm U/2$ and thus
represents an insulator. For $V\ll U$, $\epsilon_1\simeq6V^2/U$,
$\epsilon_2\simeq U/2+10V^2/U$, and $a_1\simeq18V^2/U^2$. The
appearance of spectral weight at $\pm\epsilon_1$, close to the ``Fermi
level'' at zero energy, is the best approximation to a metallic phase
which is possible within our model. In the limit $V\rightarrow0$, this
spectral weight disappears, thus simulating a ``metal-to-insulator
transition.'' If we think of the two $\delta$ function contributions
at $\pm\epsilon_1$ to the single-particle spectrum as representing the
quasiparticle resonance (QPR), its total weight is given by
$Z\equiv2a_1\simeq36V^2/U^2$ and its width by
$T^*\equiv2\epsilon_1\simeq ZU/3$.

From Eqs.\ (\ref{green})-(\ref{a1}), we obtain the noninteracting 
Green's function $G_0$ and the self-energy $\Sigma(z)\equiv
G_0^{-1}(z)-G^{-1}(z)$, 
\begin{eqnarray}
 G_0(z)&=&\frac{1}{2}\left(\frac{1}{z-V}+\frac{1}{z+V}\right)
	=\frac{z}{z^2-V^2},
\label{green0}\\
 \Sigma(z)&=&\frac{U^2}{8}\left(\frac{1}{z-3V}+\frac{1}{z+3V}\right)
	=\frac{U^2}{4}\frac{z}{z^2-9V^2}.
\label{self-energy}
\end{eqnarray}
The first quantity is completely determined by $V^2$, while the second
one also depends on $U^2$. The poles of the self-energy at $\pm3V$
are precursors of the $1/z$ pole of the insulating phase. Note, that
close to the transition point, $V\ll U$, these poles are located in
the gap of the single-particle spectrum between the QPR and the
``Hubbard bands''. The reason is that, while the width $T^*$ of the
QPR collapses linearly in $ZU$, the self-energy poles move more slowly
towards the Fermi level at $\omega=0$, according to a $\sqrt{Z}U$
behavior. Note, that at least the dependences on $Z$ are in agreement
with the scenario of Ref.\ \cite{Zhang:1993} on the Mott transition in
the Hubbard model: There, $T^*\sim ZD$, while the self-energy acquires
poles at about $\pm \sqrt{Z}\,D$ ($D$ is the half bandwidth of the
bare band).

We now consider the self-energy as a functional of the
noninteracting Green's function. Eqs.\ (\ref{green0}) and 
(\ref{self-energy}) imply
\begin{equation}
 i\Sigma(t)=U^2[iG_0(t)]^3,
\label{IPT}
\end{equation}
which means that the IPT approximation \cite{Georges:1996} becomes
exact in our toy model.

Next, we ask what we would obtain in leading-order self-consistent
perturbation theory. The leading-order skeleton diagram for the
self-energy is equivalent to Eq.\ (\ref{IPT}), except that we have to
replace the noninteracting Green's function by the fully renormalized
one of Eq.\ (\ref{green}). In the frequency domain, we obtain
\begin{eqnarray}
 \Sigma(z)&=&U^2\left(
	a_1^3\left[\frac{1}{z-3\epsilon_1}
             +\frac{1}{z+3\epsilon_1}\right]
\right.\nonumber\\
  &&+3a_1^2a_2\left[\frac{1}{z-(2\epsilon_1+\epsilon_2)}
             +\frac{1}{z+(2\epsilon_1+\epsilon_2)}\right]
\nonumber\\
  &&+3a_1a_2^2\left[\frac{1}{z-(\epsilon_1+2\epsilon_2)}
             +\frac{1}{z+(\epsilon_1+2\epsilon_2)}\right]
\nonumber\\
  &&\quad\;+a_2^3\left.\left[\frac{1}{z-3\epsilon_2}
             +\frac{1}{z+3\epsilon_2}\right]\right).
\label{2nd-selfconsistent-pt}
\end{eqnarray}
As we approach the ``insulating phase,'' $V\rightarrow0$, only poles 
at $\pm3U/2$ with residues $U^2/8$ survive. This approximation,
therefore, fails to bring about the $1/z$ pole characteristic of the
insulating phase. Its failure to account for the insulating phase
becomes even more evident if we translate the result of Eq.\
(\ref{2nd-selfconsistent-pt}) into a single-particle spectral
function. For $V=0$, we find a spurious $\delta$ function contribution
at the Fermi level, that carries 90\% of the spectral weight. This
problem is not specific to our toy model. In the context of the
infinite-dimensional Hubbard model, second-order self-consistent
perturbation theory suffered from exactly the same flaws
\cite{Mueller-Hartmann:1989b}. There, too, it always predicts the
existence of a QPR with a spectral weight that tends to be too large.  

Finally, we express the exact self-energy as a functional of the full
Green's function (\ref{green}) with the goal of elucidating why its
expansion in powers of $U$ might break down as we approach the
``transition point,'' $V\rightarrow0$. The skeleton expansion of the
self-energy is obtained by regarding $G$ as a functional of $G_0$ and
$U$, $G[G_0,U]$, and by substituting the inverted functional
$G_0[G,U]$ into $\Sigma[G_0,U]$. This requires the knowledge of $G$
for an arbitrary $G_0$ and thus the solution of the most general
impurity model, which is hopeless. Nevertheless, the essential element
of the skeleton expansion is that the self-energy is viewed as a
functional of $G$ and $U$ rather than $G_0$ and $U$. Since in our
case, $G_0$ is parametrized by a single parameter, $V^2$,
straightforward perturbation theory consists in expanding
$\Sigma(V^2,U)$ in powers of $U$. By contrast, renormalized
perturbation theory corresponds to first trading $V^2$, which
completely determines $G_0$, for one of the  three parameters that
determine the Green's function (\ref{green}), say, $\epsilon_1$, and
afterwards expanding $\Sigma(\epsilon_1 ,U)$ in powers of $U$. The
procedure of going over from $\Sigma(V^2,U)$ to $\Sigma(\epsilon_1,U)$
corresponds to performing a Legendre transformation. Each of the Eqs.\
(\ref{eps1})-(\ref{a1}) can be solved uniquely for $V^2$:
\begin{eqnarray}
 V^2&=&\frac{\epsilon_1}{18}\left(10\epsilon_1
	+\sqrt{64\epsilon_1^2+9U^2}\right),
\label{skel1}\\
 V^2&=&\frac{\epsilon_2}{18}\left(10\epsilon_2
	-\sqrt{64\epsilon_2^2+9U^2}\right),
\label{skel2}\\
 V^2&=&\frac{9-40a_1+80a_1^2+3(4a_1-1)\sqrt{9-8a_1+16a_1^2}}
	{1024a_1(1-2a_1)}\,U^2
\label{skel3}.
\end{eqnarray}
Depending on which of these equations we insert into Eq.\
(\ref{self-energy}), we obtain a different parametrization of
$\Sigma[G,U]$. If renormalized perturbation theory were well-behaved
all the way down to the transition point, $V\rightarrow0$, all
parametrizations of $\Sigma[G,U]$ would also be well-behaved when
expanded in powers of $U$. Yet, the expansion of
$\Sigma(\epsilon_1,U)$ in powers of $U$ involves an expansion of the
square root in Eq.\ (\ref{skel1}) in powers of $U/\epsilon_1$, which
converges only for $U<8\epsilon_1/3$. But since
$\epsilon_1\rightarrow0$ as $V\rightarrow0$, this expansion has zero
radius of convergence as one approaches the point $V=0$, which is the
analog of the Mott transition. More specifically, this expansion fails
to converge for $V^2/U^2\le(5+4\sqrt{2})/64\simeq0.167$, or for
$Z\le9(4-\sqrt{2})/28\simeq83\%$. This result indicates that the
skeleton expansion of the self-energy is increasingly misbehaved as
the Mott transition is approached, even though the actual range of
convergence seems to be widely underestimated by our toy model.

In summary, I have demonstrated within a simple toy model that
neither second-order self-consistent perturbation theory nor a
skeleton expansion of the self-energy to all orders in $U$ can account
for the Mott transition properly. While the former approach always
predicts a metallic phase and overestimates the spectral weight of the
quasiparticle resonance, the latter one fails to converge in the
vicinity of the transition. By contrast, ordinary leading-order
perturbation theory in $U$ is well-behaved and, in our simple toy
model, even turns out to be exact.

I thank G.\ Kotliar for many stimulating discussions and the friendly
working environment. I would also like to thank the Deutsche  
Forschungsgemeinschaft for financial support.


\begin{thebibliography}{10}

\bibitem{Mott:1961}
N.~F. Mott, Philos. Mag. {\bf 6},  287  (1961).

\bibitem{Hubbard:1963}
J. Hubbard, Proc. Roy. Soc. London A {\bf 276},  238  (1963).

\bibitem{Georges:1996}
A. Georges, G. Kotliar, W. Krauth, and M.~J. Rozenberg, Rev. Mod. Phys. {\bf
  68},  13  (1996).

\bibitem{Metzner:1989a}
W. Metzner and D. Vollhardt, Phys. Rev. Lett. {\bf 62},  324  (1989).

\bibitem{Zhang:1993}
X.~Y. Zhang, M.~J. Rozenberg, and G. Kotliar, Phys. Rev. Lett. {\bf 70},  1666
  (1993).

\bibitem{Moeller:1995}
G. Moeller {\it et~al.}, Phys. Rev. Lett. {\bf 74},  2082  (1995).

\bibitem{Baym:1961}
G. Baym and L. Kadanoff, Phys. Rev. {\bf 124},  287  (1961).

\bibitem{Mueller-Hartmann:1989b}
E. M\"uller-Hartmann, Z. Phys. B {\bf 76},  211  (1989).

\bibitem{Menge:1991}
B. Menge and E. M\"uller-Hartmann, Z. Phys. B {\bf 82},  237  (1991).

\bibitem{Georges:1992a}
A. Georges and G. Kotliar, Phys. Rev. B {\bf 45},  6479  (1992).

\bibitem{Yosida:1970}
K. Yosida and K. Yamada, Prog. Theor. Phys. {\bf 46},  244  (1970).

\bibitem{Yosida:1975}
K. Yosida and K. Yamada, Prog. Theor. Phys. {\bf 53},  1286  (1975).

\bibitem{Yamada:1975}
K. Yamada, Prog. Theor. Phys. {\bf 53},  970  (1975).

\bibitem{Kajueter:1996b}
H. Kajueter and G. Kotliar, Phys. Rev. Lett. {\bf 77},  131  (1996).

\end{thebibliography}
\end{document}